\documentclass[12pt]{iopart}

\usepackage{iopams}
\usepackage{graphicx}
\usepackage{bm}
\begin{document}

\title{Perturbations of global monopoles as a black hole's hair}

\author{Hiroshi Watabe and Takashi Torii}
\address{Advanced Research Institute for Science and Engineering,Waseda University, Shinjuku-ku, Tokyo 169-8555, Japan}

\eads{\mailto{watabe@gravity.phys.waseda.ac.jp}, \mailto{torii@gravity.phys.waseda.ac.jp}}

\begin{abstract}
We study the stability of a spherically symmetric black hole with a global monopole hair.
Asymptotically the spacetime is  flat but has a deficit solid angle which depends on the vacuum
expectation value of the scalar field. When the vacuum  expectation value is larger than a certain
critical value, this spacetime has a cosmological event horizon. We investigate the stability of these
solutions against the spherical and polar perturbations and confirm that the global monopole hair is stable
in both cases. Although we consider some particular modes in the polar case, our analysis 
suggests the conservation of the ``topological charge" in the presence of the event horizons and
violation of black hole no-hair conjecture in asymptotically non-flat spacetime.\end{abstract}

%<<<<<<<<<<<<< PACS NUMBER >>>>>>>>>>>>>>>%
%\pacs{04.70.Bw, 04.20.Jb, 11.27.+d, 04.40.-b}

% 04.70.Bw Classical black holes
% 04.20.Jb Exact solutions 
% 11.27.+d Extended classical solutions; cosmic strings,
% domain walls, texture (see also 98.80.Cq in cosmology)
% 04.40.-b Self-gravitating systems; continuous media and classical fields in curved spacetime

%98.80.Cq Particle-theory and field-theory models of the early Universe (including cosmic pancakes, cosmic strings, chaotic phenomena, inflationary universe, etc.) (see also 11.25.-w Strings and branes, and 11.10.-z in general theory of fields and particles) 

%<<<<<<<<<<<<< PREPRINT NUMBER >>>>>>>>>>>>>>>%
%\hspace{35mm} 
%gr-qc/030x0xx

\maketitle

%======================================%
%<<<<<<<<<<<< SECTION I  >>>>>>>>>>>>>>%
%======================================%
\section{Introduction}
The exterior gravitational field of a stationary source may have a large number of independent
multipole moments. However, if a black hole event horizon (BEH) is formed, these multipole
moments will reduce to three physical parameters, $M$, $a$ and $Q$. These parameters are
interpreted as mass, angular momentum and net electromagnetic charge of the black hole, respectively.
This statement is called the black hole no-hair conjecture, which was proposed by Ruffini
and Wheeler\cite{non-hair}.

Many candidates for counter-examples were proposed to investigate whether this conjecture is true or not.
Among them the colored black hole in Einstein-Yang-Mills system is interesting\cite{EYM}.
Although this solution is unstable\cite{EYMunstable},
it opened up a new possibility that the black hole may have various matter hairs\cite{To1}.
Actually stable non-Abelian black hole solutions were found as the
black hole counter-part of the self-gravitating topological defects and solitons\cite{skyrme,mono}.

Another possibility for the hairy black hole is the solution in asymptotically non-flat spacetime.
One of us showed that the black hole can support a scalar hair in asymptotically
de Sitter\cite{To2} and anti-de Sitter spacetime\cite{torii}. 
The colored black hole in anti-de Sitter spacetime was also reported\cite{Win}.
Some of these solutions were found to be stable, so they can be  strong
counter-examples to the black hole no-hair conjecture.

Phase transitions in the early universe are caused by  symmetry breaking
leading to a manifold of degenerate vacua with nontrivial topology and
giving rise to topological 
defects.  The topological defects are classified  into
domain walls, cosmic strings and monopoles by the topology of the vacua.
If the gauge field is involved in the spontaneous symmetry breaking,
the topological defects are gauged. On the other hand, when the symmetry
is global, the emerging defects are called global defects.

Although energy of  the gauge 
monopole is finite,  the global monopole has divergent energy
because of the long tail of the scalar field.
This divergence has to be removed by cutting it off at a certain distance. 
This procedure is not necessarily artificial in the early universe, because other defects
which may exist near the original one cancel the divergence. 
These neighboring defects are not only the monopoles, but also can be 
domain walls or cosmic strings.

When general relativistic self-gravity is considered, such a divergent behavior can be made to disappear
by a new definition of the energy\cite{newADM}.
Spacetime becomes asymptotically flat but has a deficit solid angle\cite{Barriola}.
Moreover, the motion of a test particle around the global monopole perceives repulsive force 
from the center\cite{Harari}.

This kind of global monopole has a different asymptotic behavior from 
the non-gravitating one\cite{Liebling}.
As the vacuum expectation value (VEV) $v$ of the
$O(3)$ scalar field increases, the deficit solid angle also 
gets large and  becomes $4\pi$ when $v =v_{cri}:=\sqrt{1/8\pi}\approx0.199$.
Beyond this critical value there is no ordinary monopole solution, but
a new type of solution appears in the parameter range 
$v_{cri} < v < v_{max} :=\sqrt{3/8\pi}\approx0.345$.
This has a cosmological event horizon (CEH) at $r=r_c$. We call this the  supermassive global monopole
in imitation of the supermassive global strings\cite{string}. 
When $v>v_{max}$, there are only trivial de Sitter solutions under
the hedgehog ansatz. This solution has a configuration where
the scalar field sits on the top of the potential barrier.

One of the most important issues for these kinds of isolated objects
is  stability.  In the non-gravitating monopole case,
if Derrick's no-go theorem\cite{Derrick} could be applied,
they would be unstable towards a radial rescaling
of the field configuration. This is, however, not the case due to
the diverging energy of the solutions. It was demonstrated that the
 monopole solutions are stable against spherical perturbations.
As for the non-spherical perturbations, several studies have been done.
Consequently, the non-gravitating global monopole is stable against both
spherical and polar perturbations. That is consistent with the conservation
of the topological charge.

In the self-gravitating case, the global monopole can have a CEH.
Thus we can not define exact topological charges because of the peculiar asymptotic structure.
One may find an exact proof of the conservation of the ``topological charges" defined
 by the analogy
 even in the spacetime
with the BEH and/or CEH by putting the in-going and/or out-going conditions
at each horizon. This issue is very interesting but beyond the
scope of the current paper.

Maison and Liebling studied the stability of the static solution against spherical
and hedgehog type perturbations\cite{Maison}.
They found that the supermassive monopole and de Sitter solutions are 
stable when $v<v_{max}$. We investigated the stability against polar
perturbations\cite{watabe},
without the spherical symmetry nor the hedgehog ansatz.
We obtained the conclusion that the (supermassive) monopole solution is stable while 
the de Sitter solution is always unstable.

If a black hole collides with and absorbs  a global monopole, the black hole would have
$O(3)$ scalar hair. This type of  solution was found  numerically\cite{Liebling}. 
The spacetime of this solution is asymptotically flat but with a deficit solid angle
or has the CEH.
There emerges a question of whether the black hole no-hair conjecture holds
in such a spacetime. This is the issue of this paper.
 
This paper is organized as follows. 
In Sec.~\ref{sec:static}, we introduce the model and review the global monopole and the 
black hole solutions.
In Sec.~\ref{sec:spherical}, we analyze the spherical perturbation of the solutions.
In Sec.~\ref{sec:deform}, we formulate the polar perturbations of the $O(3)$ scalar field and the metric, and
show the stability 
of the black hole with global monopole hair. 
Throughout this  paper, we use the units $\hbar=c=G=1$.

%======================================%
%<<<<<<<<<<<< SECTION II  >>>>>>>>>>>>>%
%======================================%
\section{Black hole solution with the global monopole hair}
\label{sec:static}
In this section, we briefly review the self-gravitating
global monopole solution\cite{Barriola,Maison} and
its black hole counter-part.
We consider  a scalar field which has spontaneously broken internal
$O(3)$ symmetry, and minimally couples to gravity.
The action is

%===========<Equation>============%
\begin{equation}
S=\int d^4x \sqrt{-g}
\left[\frac{R}{16 \pi}
-\frac{1}{2}\partial_{\mu}\Phi^{a} \partial^{\mu} \Phi^{a}
-\frac{\lambda}{4}\bigl(\Phi^2 -v^2\bigr)^2 \right],
\end{equation}
%=================================%
where $R$ is the Ricci scalar of the spacetime and $\Phi^{a}$ $(a=1,2,3)$
is the triplet scalar field.
$\lambda$ and $v$ are the self-coupling constant and
the VEV of the scalar field, respectively.
The energy momentum tensor is 
%===========<Equation>============%
\begin{equation}
T_{\mu \nu}
=\partial_{\mu}\Phi^{a} \partial_{\nu}\Phi^{a}
-g_{\mu\nu}\left[\frac{1}{2}\partial_{\rho}\Phi^{a}\partial^{\rho}\Phi^{a}
+\frac{\lambda}{4}\bigl(\Phi^2-v^2\bigr)^2 \right].
\end{equation}
%=================================%
For a static solution with unit winding number, we adopt the so-called hedgehog ansatz to the scalar field,
%===========<Equation>============%
\begin{equation}
\Phi^{a}=h(t,r)\frac{x^a}{r},
\end{equation}
%=================================%
where $x^a$ are the Cartesian coordinates.
We shall consider spherically symmetric static spacetime,
%===========<Equation>============%
\begin{equation}
ds^2=-f(r)e^{-2\delta(r)}dt^2 +\frac{1}{f(r)}dr^2
+r^2 (d \theta^2 +\sin^2\theta d \phi^2),
\end{equation}
%=================================%
where 
%===========<Equation>============%
\begin{equation}
f(r)=1-\frac{2m(r)}{r}.
\end{equation}
%=================================%
By  scaling the variables as
%===========<Equation>============%
\begin{equation}
\bar{x}^{\mu}=v\sqrt{\lambda}x^{\mu},
\,\,\,\, 
\bar{m}=v\sqrt{\lambda}m,
\,\,\,\,
\bar{h}=\frac{h}{v},
\,\,\,\,
\bar{\Phi}=\frac{\Phi}{v},
\end{equation}
%=================================%
the action can be rewritten as
%===========<Equation>============%
\begin{equation}
S=\int d^4x \sqrt{-g}\left[
\frac{\bar{R}}{16 \pi v^2}
-\frac{1}{2}\partial_{\mu}\bar{\Phi}^{a} \partial^{\mu} \bar{\Phi}^{a}
-\frac{1}{4}\bigl(\bar{\Phi}^2-1\bigr)^2 \right].
\end{equation}
%=================================%
In this formula, the coupling constant $\lambda$ is scaled out. The VEV appears only in
the denominator of the curvature term and affects the system only when self-gravity is
taken into account. 
The basic equations are
%===========<Equation>============%
\begin{equation}
\frac{d m}{d r}
=4 \pi v^2
r^2\left[\frac{1}{2}fh^{\prime 2}+\frac{h^2}{r^2}+\frac{1}{4}(h^2-1)^2\right],
\end{equation}
%=================================%
\begin{equation}
\frac{d \delta}{d r}=-4 \pi v^2 r {h^{\prime}}^2,
\end{equation}
%=================================%
\begin{equation}
\bigl(r^2e^{-\delta}fh'\bigr)'-2he^{-\delta}=r^2e^{-\delta}(h^2-1)h,
\label{eqn:h0}
\end{equation}
%=================================%
where a prime denotes a derivative with respect to the radial coordinate.
We have omitted the bar of the variables.

We integrate these equations with suitable boundary conditions. If we put the regularity condition at
the center, we will obtain the self-gravitating monopole solution. 
In the black hole case the spacetime and the scalar field should be regular at the BEH. By
this condition we find
$h$ at BEH ($r=r_b$) can be regarded as a free parameter and  the other variables are determined by
$h(r_{b})$. We choose $\delta(r_b)=0$. Then $\delta$ approaches some constant value $\delta_{\infty}$
as $r\to \infty$. The ordinary asymptotic metric is recovered by rescaling the time
coordinate as $t:=te^{-\delta_{\infty}}$.
The free parameter $h(r_b)$ is determined by the other boundary conditions at
$r\rightarrow \infty$ for the spacetime without the CEH or at the CEH
($r=r_{c}$) for the supermassive case.

\begin{figure}[tb]
\centerline{
\includegraphics[height=7cm]{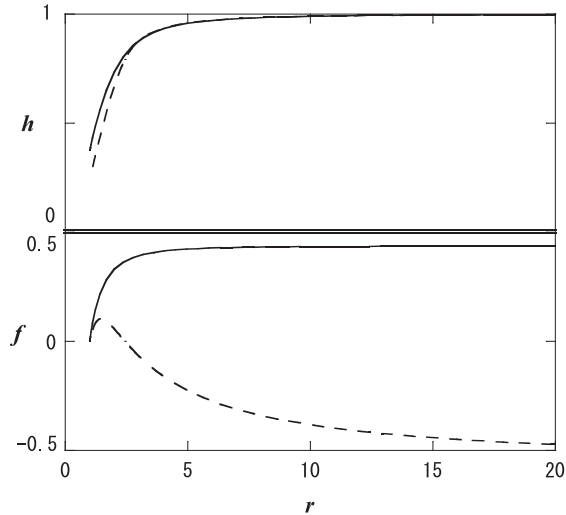}
}
\caption{The configurations of the scalar field  $h$ and metric function $f$ of the monopole black hole for 
$v=0.15<v_{cri}$ (solid line: the ordinary case) and $v=0.25>v_{cri}$ (dashed line: the supermassive case). 
We fix $r_{b}=1.0$.
In the ordinary case, 
the metric function $f$ approaches a
constant value less than $1$ as $r \rightarrow \infty$. This means that the spacetime has a
deficit solid angle.
For the supermassive case, the metric function $f$ crosses zero at the CEH.
}
\label{fig:static}
\end{figure}

Whether the self-gravitating monopole and the monopole black hole have the CEH or not depends on
the VEV. When
$0<v<v_{cri}$, the spacetime does not have the CEH. The configuration of the scalar
field and the metric are shown in Fig.~\ref{fig:static} (solid lines). 
For the case in which the VEV is larger than the critical value $v_{cri}$, the spacetime has the CEH. 
In Fig.~\ref{fig:static}  we can also see this feature (dashed line). 
There is the maximum value $v_{max}$
where the supermassive global monopole (BH) solution coincides
with the (Schwarzschild-)de Sitter solution. Beyond 
$v_{max}$ no non-trivial solution exists.

For the regular monopole case without the BEH Maison and Liebling investigated the stability against
the spherical (both in spacetime and in internal space) perturbations. They found that both the
ordinary and the supermassive monopole solutions are stable. We extended their analysis and
investigated the stability against the polar perturbations. We found that both of monopole solutions
are stable while the de Sitter solution is always unstable\cite{watabe}.

%-------------<fig 1>-------------------------
%---------------------------------------------

%--------------------<fig2>-------------------------------------------------------
\begin{figure}[tb]
\centerline{
\includegraphics[height=7cm]{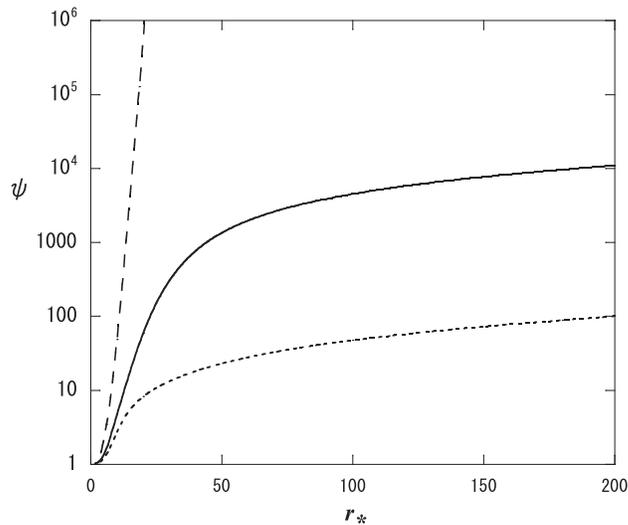}
}
\caption{The configurations of the integrated function for $v=0.15$
(dashed line), 
$v=0.25$ (solid line) and $v=0.28$ (doted line) with $\sigma^2=0$. They
do not have a node but are positive definite outside of the BEH.
This indicates nonexistence of the unstable modes. }
\label{fig:spherical}
\end{figure}
%-------------------------------------------------------------------------------

%======================================%
%<<<<<<<<<<<< SECTION III  >>>>>>>>>>>>%
%======================================%
\section{Spherical perturbation of the global monopole hair}
\label{sec:spherical}

In the remainder of this paper, we investigate the stability of the black hole solutions with the
ordinary and the supermassive global monopole hair. Since our solutions have event horizons,
the ``topological charge" may not be conserved. So our analysis is interesting from the view point of
the defects in spacetime with horizons. This issue is also interesting in relation to the black hole
hair.

The
metric perturbations which we adopt here are,
%===========<Equation>============%
\begin{eqnarray}
&& m(r) \rightarrow m(r)+\tilde{m}(r)e^{i\sigma t},
\\
&& \delta(r) \rightarrow \delta(r)+\tilde{\delta}(r) e^{i\sigma t}.
\end{eqnarray}
%=================================%
The the perturbation of the scalar field is
%===========<Equation>============%
\begin{eqnarray}
h(r)\rightarrow h(r)+\tilde{h}(r)e^{i\sigma t}.
\end{eqnarray}
%=================================%
We introduce new variable $\psi$,
%===========<Equation>============%
\begin{eqnarray}
\tilde{h}(r)=\frac{\psi(r)}{r}. \nonumber
\end{eqnarray}
%=================================%
Then the perturbation equations becomes a Schr\"odinger type equation as follows,
%===========<Equation>============%
\begin{equation}
-\frac{d^2\psi}{dr_{*}^2}+U(r)\psi=\sigma^2\psi,
\label{eqn:spherical}
\end{equation}
%=================================%
where $r_{*}$ is the tortoise coordinate ($d/dr_{*}=fe^{-\delta}d/dr$) and
%===========<Equation>============%
\begin{eqnarray}
U(r)=e^{2\delta}f\left[\frac{f'}{r}
-4\pi v^2fh^{\prime 2}\biggl(1+2r\frac{f'}{f}+8\pi v^2h^2\biggr)\right.
\nonumber
\\
\left.+\frac{2}{r^2}+16\pi r v^2\biggl(\frac{2}{r^2}+h^2-1\biggr)hh'+3h^2-1\right].
\end{eqnarray}
%=================================%

As the boundary conditions of the perturbation functions, we assume the in-going boundary
conditions at the BEH. Although, one usually puts out-going boundary condition at infinity to
calculate quasi-normal modes of an isolated object, we leave it free as long as the perturbation
functions do not diverge because we are not interested in the definite quasi-normal modes but
in the stability.
If we put the out-going boundary condition, which is more restricted condition than
ours, and there
are unstable modes, these modes are also found out by our analysis.

What we will do is search the unstable mode by integrating Eq.~(\ref{eqn:spherical}) for
various value of $\sigma$. As a result, we could not find any unstable modes similar to the 
self-gravitating monopole analyzed by Maison and Liebling\cite{Maison}.

Although our numerical
analysis does not give exact proof of the non-existence of the unstable modes, we can
show a basis for stability.  

Fig.~\ref{fig:spherical} is the configurations of the integrated function with $\sigma^2=0$.
If this function has at least one node, we would find the unstable modes.
However, the function with any $v$ does not have a node but is positive definite outside
of the BEH. Thus we may conclude that the static solutions are stable against spherical
perturbations.

%======================================%
%<<<<<<<<<<<< SECTION IV  >>>>>>>>>>>>>%
%======================================%
\section{Polar perturbation of the global monopole hair}
\label{sec:deform}

Several studies have been made on the non-spherical perturbations of the non-gravitating global
monopole\cite{Goldhaber,Bennett,Achucarro}. 
Goldhaber proposed the polar deformation of the $O(3)$ scalar field,
%===========<Equation>============%
\begin{eqnarray}
\Phi^{1}&=&H(t,r,\theta)\sin \tilde{\theta}(t,r,\theta) \cos \phi,  
\nonumber \\
\Phi^{2}&=&H(t,r,\theta)\sin \tilde{\theta}(t,r,\theta) \sin \phi, 
\\
\Phi^{3}&=&H(t,r,\theta)\cos \tilde{\theta} (t,r,\theta), 
\nonumber
\end{eqnarray}
%=================================%
where
%===========<Equation>============%
\begin{eqnarray}
\tan\biggl(\frac{\tilde{\theta}}{2}\biggr)=e^{\Theta+\delta\Theta(t,r,\theta)},
\end{eqnarray}
%=================================%
$\tilde{\theta}$ is the polar component of $\Phi^a$, $\Theta\!:=\ln[\tan(\theta/2)]$ and 
$H(t,r,\theta)=h(r)+\delta h(t,r,\theta)$.
When $\delta\Theta\equiv 0$, i.e., $\tilde{\theta}\equiv\theta$ and 
$\delta h \equiv \delta h(t, r)$, the global 
monopole is spherically symmetric, while it become a ``string'' along  the $z$-axis
when $\delta\Theta \rightarrow \infty$.
The energy of the static global monopole is expressed with the new coordinate 
$\Theta$,
%===========<Equation>============%
\begin{eqnarray}
E&=&\int dr d\Theta d\phi (\rho_1+\rho_2),
\end{eqnarray}
%=================================%
where
%===========<Equation>============%
\begin{eqnarray}
\rho_1&=&\frac{H^2}{2}\biggl[\sin^2\tilde{\theta}
+\biggl(\frac{\partial \tilde{\theta}}{\partial \Theta}\biggr)^2 
+\frac{r^2}{\cosh^2 \Theta}
\biggl(\frac{\partial \tilde{\theta}}{\partial r} \biggr)^2\biggr],
\nonumber\\
\rho_2&=&\frac{1}{2}\biggl(\frac{\partial H}{\partial \Theta}\biggr)^2
+\frac{r^2}{2\cosh^2 \Theta}
\biggl[
\biggl(\frac{\partial H}{\partial r}\biggr)^2
+\frac{1}{2}(H^2-1)^2\biggr].
\end{eqnarray}
%=================================%
In the far region from the monopole core, the terms  
$\partial H/\partial r$ and $\partial H/\partial \Theta$ can be neglected. 
Similarly one can neglect $\partial \tilde{\theta}/\partial r$ and $(H^2-1)$ 
(See Ref.~\cite{Goldhaber} for details).
Goldhaber pointed out that the remaining
first two terms in $\rho_1$ have the same form as the 
energy of a sine-Gordon soliton, 
which implies that the energy is invariant under translation of 
coordinate $\Theta$ (or $\xi$).

We adopt Goldhaber's formula to investigate the stability of the black hole with a global monopole
hair.
%We consider the $l=1$ Legendre type perturbations for the scalar field as,
We consider the  simplest case,
%===========<Equation>============%
\begin{eqnarray}
\Phi^{1}&=&\Bigl[h+(\eta +h\xi)\cos \theta e^{i \sigma t}\Bigr] 
\sin\theta \cos\phi,  
\nonumber \\
\Phi^{2}&=&\Bigl[h+(\eta +h\xi)\cos \theta e^{i \sigma t}\Bigr]
\sin\theta \sin\phi, 
\\
\Phi^{3}&=&\Bigl[h+(\eta +h\xi)\cos \theta e^{i \sigma t}\Bigr]
\cos \theta -h\xi e^{i \sigma t}  
\nonumber,
\label{eqn:ptb}
\end{eqnarray}
%=================================%
where $\delta\Theta=\xi(r) e^{i\sigma t}$ and 
$\delta h=\eta(r) \cos \theta e^{i\sigma t}$.
This corresponds to the polar perturbation with $l=1$.
In this paper we consider only the $l=1$ case because the  perturbation corresponding to  $l \geq 2$ 
of Goldhaber's formula are so complicated that explicit forms of them, like orthogonal functions,
have  not been known.

Since the matter field of the background solution is non-zero, the first order
perturbations of the matter field produce the metric perturbations at the first order level, which can be
described as
%===========<Equation>============%
\begin{eqnarray}
ds^2&=&-fe^{-2\delta}e^{\delta \nu}dt^2
+\frac{1}{f}e^{\delta \mu_{2}} dr^2
+r^2e^{\delta \mu_{3}}d \theta^2 
+r^2 \sin^2 \theta e^{\delta \psi}d\phi^2.
\end{eqnarray} 
%=================================%
The first order variables in the metric can be separated \cite{Friedman},
%===========<Equation>============%
\begin{eqnarray}
\delta \nu &=&\sum N_{l}(t,r)P_{l}(\cos \theta), 
\nonumber\\
\delta \mu_{2}&=&\sum L_{l}(t,r) P_{l}(\cos \theta), 
\nonumber\\
\delta \mu_{3}&=&\sum [T_{l}(t,r)P_{l}(\cos \theta) 
+S_{l}(t,r) P_{l,\theta,\theta}(\cos \theta)], 
\nonumber \\
\delta \psi&=&\sum [T_{l}(t,r) P_{l}(\cos \theta) 
+S_{l}(t,r)P_{l,\theta}(\cos \theta) \cot \theta],
\end{eqnarray}
%=================================%
where $P_{l}$ is a Legendre polynomial.
Considering the simplest case $l=1$, which is consistent with $l=1$ Goldhaber's formula, we drop the suffix $l$. 
Thus, the perturbed 
metric can be written in the form,
%===========<Equation>============%
\begin{eqnarray}
ds^2&=&-\left[f+B(r)e^{i \sigma t} \cos \theta\right]
e^{-2\delta}dt^2 
+\frac{1}{f}\left[1+L(r)e^{i \sigma t} \cos \theta\right]dr^2 \nonumber \\
& &+r^2\left[1+T(r)e^{i \sigma t}\cos \theta\right]
(d \theta^2+\sin^2 \theta d \phi^2).
\label{eqn:met-ptb}
\end{eqnarray}
%=================================%
Here, we have introduced a new variable $B(r):=N(r)f(r)$ for convenience,
re-defined $T := T-S$ and assumed harmonic time dependence.

We can get the perturbation equations
%===========<Equation>============%
\begin{equation}
L+T=16 \pi v^2h^2 \xi,
\label{eqn:Ttheta}
\end{equation}
%=================================%
\begin{equation}
-(rT)'+r\biggl(\frac{f'}{2f}-\delta'\biggr)T+L=8 \pi v^2 rh'\eta,
\label{eqn:TR}
\end{equation}
%=================================%

\begin{eqnarray}
&&\frac{\sigma^2e^{2\delta}r^2 T}{f}
-f\biggl(1+\frac{rf'}{f}-2r\delta'\biggr)L
-\frac{1}{f}(1+rf')B+rB'
%\nonumber \\ 
%&& ~~~~~~~~~~
+rf\biggl(1+\frac{rf'}{2f}-r\delta'\biggr)T' 
\nonumber \\
&&~~~~~~~~~
=8\pi v^2 \biggl[ fr^2h'\eta'+h^2T-2h(\eta+h\xi) 
%\nonumber \\
%&&~~~~~~~~~~
-r^2(h^2-1)h\eta-\frac{1}{2}fr^2{h'}^{2}L \biggr],
\label{eqn:RR}
\end{eqnarray}
%=================================%
from the Einstein equations and
%===========<Equation>============%
\begin{eqnarray}
&& \frac{\sigma^2  e^{2\delta}h\xi}{f}+f(h\xi)''
+f\left(\frac{2}{r}
+\frac{f'}{f}-\delta'\right)(h\xi)'
\nonumber \\
&& ~~~~~
-\frac{2\eta+2h\xi}{r^2}+\frac{hL}{2r^2}
-\frac{hB}{2fr^2}
-h(h^2-1)\xi=0,
\label{eqn:xi}
\end{eqnarray}
%=================================%
\begin{eqnarray}
&& \frac{\sigma^2 e^{2\delta}}{f}\eta+f\eta''
+f\left(\frac{2}{r}
+\frac{f'}{f}-\delta'\right) \eta' +fh'T'
%\nonumber \\
%&& \;\;\;\;
-\left[\frac{2h}{r^2}
+ h(h^2-1) \right]L
\nonumber \\
&&~~~~~~
-\frac{4h\xi+4\eta-2hT}{r^2}
-\frac{fh'L'}{2}+\frac{h'B'}{2}-\frac{h'f'B}{2f}
-(3h^2-1)\eta
=0,
\label{eqn:eta}
\end{eqnarray}
%=================================%
from the scalar field equation.

We assume the behavior of the variables near $r=r_b$ to satisfy the regularity
condition as,
%===========<Equation>============%
\begin{equation}
A=A_0+A_{1}\rho+\frac12 A_{2}\rho^2+\dots,
\label{exp1}
\end{equation}
%=================================%
for the background variables such as $m$, $\delta$ and $h$, and
%===========<Equation>============%
\begin{equation}
A=\rho^\beta\biggl(A_0+A_{1}\rho+\frac12 A_{2}\rho^2+\dots\biggr),
\label{exp2}
\end{equation}
%=================================%
for the first order variables such as $B$, $L$, $T$, $\cdots$, where $\rho:=(r-r_b)/r_b$.

Expanding the first order equations (\ref{eqn:Ttheta})-(\ref{eqn:eta}) with the Eqs.~(\ref{exp1}) and
(\ref{exp2}), we get, 
%===========<Equation>============%
\begin{equation}
\eta_{0}=B_{0}=T_{0}=0,
\end{equation}
%=================================%
\begin{equation}
L_{0}=16\pi v^2h_{0}\alpha_{0},
\end{equation}
%===================================%
\begin{eqnarray}
\alpha_{1}&=&\Delta
\Bigl\{ 2\bigl[(1+\delta_{1})(r_b-2m_1)+2m_{2}
\bigr]{\beta}^{2}
%\nonumber \\
%&&~~~~~~~~
+\bigl[(\delta_{1}-1)(r_b-2m_{1})+2m_{2}
\bigr]{\beta}
\nonumber \\
&&~~~~~~~~
+(r_b^2+16\pi v^2)h_0^2r_b+2r_b^2-r_b^3)
\Bigr\},
\end{eqnarray}
%=================================%
\begin{eqnarray}
\eta_{1}=4\, \Delta r_b\left( 4r_b^2h_0^2\pi v^2(h_0^2-1)+8h_0^2\pi v^2+1 \right) ,
\end{eqnarray}
%=================================%
\begin{eqnarray}
L_{1}&=&16\pi\,{v}^{2} \Delta 
\Bigl\{ 
2h_{0}\bigl[(1+\delta_{1})(r_b-2m_1)+2m_{2}
\bigr]{\beta}^{2}
\nonumber \\
&&~~~~~~
+
\bigl[(r_b-2m_{1})\bigl(2h_1-h_0(1-\delta_1)\bigr)+2h_{0}m_{2}\bigr]\beta
\nonumber \\
&&~~~~~~
+(r_b-2m_{1})h_1+h_0^3r_b(r_b^2+16\pi {v}^{2})
%\nonumber \\
%&&~~~~~~
+h_0(4m_1-r_b^3)
\Bigr\},
%\nonumber \\
\end{eqnarray}
%=================================%
\begin{equation}
B_{1}=16\pi v^2\frac{r_b-2m_{1}}{r_b}h_{0}\alpha_{0},
\end{equation}
%=================================%
\begin{equation}
T_{1}=\frac{32\pi v^2 h_{0}\alpha_{0}}{1+2\beta},
\end{equation}
%=================================%
\begin{equation}
\beta^2=-\frac{r_b^2 \sigma^2}{(1-\frac{2m_1}{r_b})^2},
\end{equation}
%=================================%
where $\alpha:=h\xi$ and
%=================================%
\begin{equation}
\Delta:=\frac{\alpha_{0}}{(2\beta+1)(r_b-2m_{1})}.
\end{equation}
%=================================%
Since now $A_i$ of background fields are obtained from the static solution,
these equations are relations between the expansion coefficients $A_i$ of the perturbed  fields.
We find that all $A_i$ are specified by a value $\xi_0$. 
Hence, $\xi_0$ on the BEH are the shooting parameters 
of this equation system.
Our aim is to find an unstable mode or to indicate the non-existence of it.
The unstable mode can be assumed to have a real negative $\sigma^2$.
Therefore we assumed $\sigma^2$ 
is real and focused only on the real part of the perturbation equations.

%------------<fig3>---------------------------
%\begin{figure}[tb]
%\centerline{
%\includegraphics[height=7cm]{fig3.eps}
%}
%\caption{
%The relation of $\sigma^2_{min}$ and $r_{b}$ when $v=0.15$. The solid line is obtained by the analytical
%estimation of the asymptotic form of  the perturbation equation. Under this line 
%there is no finite mode. The dots are obtained by numerical analysis.
%We find that they are consistent with analytical one.
%}
%\label{fig:critical}
%\end{figure}

%======================================%
%<<<<<<<<<<<< SECTION V  >>>>>>>>>>>>>>%
%======================================%
\section{Results and discussion}

Having specified the boundary condition at the BEH, 
we can now integrate  the Eqs.~(\ref{eqn:Ttheta})-(\ref{eqn:eta}) numerically.
For the ordinary case, we find the critical value $\sigma^2_{min}>0$ below which there is no finite solution.
Since we have not put the strict boundary conditions at $r=\infty$ such as  
the out-going condition but just the regularity condition as in the spherical perturbation case,
these modes becomes continuous.

The existence of the critical eigenvalue can be understood from the asymptotic form of
Eq.~(\ref{eqn:eta}) at $r\rightarrow\infty$, 
%===========<Equation>============%
\begin{equation}
-\frac{d^2 \eta}{d r_{*}^2}+fe^{-2\delta}(3h^2-1)\eta=\sigma^2\eta.
\label{sch}
\end{equation}
%=================================%
$\eta$
decouples to the other first order variables and is described by a single
Schr\"odinger type equation. The 
potential $U(r):=fe^{-2\delta}(3h^2-1)\to 2fe^{-2\delta_{\infty}}$
approaches a positive value .
Hence $\sigma^2_{min}$ is determined by the asymptotic value of the potential
function as $\sigma^2_{min} \simeq 2fe^{-2\delta_{\infty}}$.

It is almost impossible to show the stability of the solution exactly by numerical
analysis because of the infiniteness of the phase space of the perturbation function.
However, the behavior of the variables with $\sigma^2=0$ gives some information about the stability. 
We find that they have no node as in the spherical case, which suggest that the solution be
stable against at least the present polar perturbations.

For the supermassive case,   the above potential term vanishes at $r=r_c$. 
Assuming the same boundary conditions as the ordinary case at $r=r_b$, we find $\sigma_{min}^2=0$
below which no finite solution exists.
We find that some of the variables can have a node when $\sigma^2=0$.
However, we also find that there is no negative $\sigma^2$ which makes all variables finite.
Moreover, if $\sigma^2$ is smaller than some negative value, all variables diverge without node.
Thus we conclude that the supermassive case is stable against the present polar perturbations, too.

We investigated the stability of black hole solutions in the Einstein-$O(3)$ scalar
system by linear perturbation. The black holes are stable against spherical perturbation
 as the particle-like case without BEH.
We also studied the polar perturbation since the trivial (Schwarzschild) de-Sitter solution
with $v_{max}<v$ is stable against the spherical perturbation\cite{Maison} but not against the non-spherical
perturbations\cite{watabe}.
We suggested that the black holes with the ordinary or the supermassive global monopole
hair
be stable against our perturbations. 
This result implies that if these kind of objects were formed in the early universe,
 they can survive without decaying to other object nor swallowing the outer matter field 
 leaving a vacuum black hole. It should be also stressed that the black hole no-hair conjecture
  is violated for $O(3)$ scalar
hair and gives new insight on the problem of black hole hair.

%======================================%
%<<<<<<<<<< Acknowledgement  >>>>>>>>>>%
%======================================%

%======================================%
%<<<<<<<<<<<< References  >>>>>>>>>>>>>%
%======================================%

\end{document}